# Observation of Saturable and Reverse Saturable Absorption at Longitudinal Surface Plasmon Resonance in Gold Nanorods

Hendry I. Elim, Jian Yang and Jim-Yang Lee*

*Department of Chemical and Biomolecular Engineering,*

*Faculty of Engineering, National University of Singapore,*

*4 Engineering Drive 4, Singapore 117576*

Jun Mi and Wei Ji*

*Department of Physics, National University of Singapore, 2 Science Drive 3,*

*Singapore 117542, Singapore*

## Abstract

Saturable and reverse saturable absorption at longitudinal surface plasmon resonance (SPR) in gold nanorods (Au NRs) have been observed using Z-scan and transient absorption techniques with femtosecond laser pulses. At lower excitation irradiances, the wavelength dispersion of saturable absorption has been determined near the longitudinal mode of SPR with a recovery time determined to be a few ten picoseconds on the SPR resonance. With higher excitation irradiances, reverse saturable absorption occurs and becomes dominant. The underlying mechanisms are discussed. Such reversible saturable absorption makes Au NRs an ideal candidate for optical limiting applications.

---

*Electronic mail: cheleejy@nus.edu.sg or phyjiwei@nus.edu.sg

In the fast evolving field of nanoscience and nanotechnology, where size and shape are crucial for the optoelectronic properties of nanomaterials, the understanding of size (or shape) dependent behavior in nanomaterials is of direct relevance to device applications and hence, intensive research efforts have been made towards this end. For example, investigations of strong enhancement of emission in gold nanorods near the surface plasmon resonance (SPR) have attracted great attention.[1-4] The SPR can be thought of as the coherent motion of the conduction-band electrons caused by interaction with an electromagnetic field.[5] Furthermore, the oscillations of the free electrons in the conduction band occupying energy states near the Fermi level give rise to a surface plasmon absorption band which depends on the length and chemical surrounding of the gold nanorods.[6] The strong absorption, scattering, and considerable local-field enhancement occurring at the SPR results in a large optical polarization associated with the collective electron oscillations.[7] Because of this, there are many interesting nonlinear optical properties of metal nanoparticles or nanorods that give immense enthusiasm for their use in applications such as optical limiting device, plasmon waveguide, sensor protection, medicine, and nanoprobes.[8-11] However, the saturable and reverse saturable absorption at longitudinal SPR in metal nanorods have not been explored and reported yet. In this letter, we report our investigations of resonant, saturable and reverse saturable absorption of longitudinal surface plasmons in gold nanorods (Au NRs). In particular, we present our observation on the wavelength dispersion of saturable absorption and dynamics of the excited electrons by using Z-scans and time-resolved transient absorption measurements, respectively.

The Au NRs were prepared according to a modification of synthesis method reported previously.[1,12-14] While details of the modification can be found elsewhere,[15] the preparation of the Au RNs is described briefly as follows. Firstly, a 2 ml aqueous solution containing 5 x $10^{-4}$ M $HAuCl_4$ and 0.1M cetyltrimethylammonium bromide (CTAB) was prepared in a conical flask. Next, 0.12 ml of ice cold 0.01 M $NaBH_4$ solution was added to the solution all at once while stirring. The solution turned pink immediately after adding $NaBH_4$, indicating particle formation. The particles in this solution were used as seeds within 2~5 hours after preparation. Secondly, 5 ml 1 x $10^{-3}$ M $HAuCl_4$ and 5 ml 0.2 M cetyltrimethylammonium bromide (CTAB) were mixed together in a clean beaker. Then, 0.25 ml 0.004M $AgNO_3$ solution and 0.065 ml of 0.0788 M fresh ascorbic acid solution were added into the solution in turn. It was found that the solution turned from orange to colorless after the addition of ascorbic acid. Finally, 0.020 ml of the seed solution was added. Within 30 minutes, the color of the solution changed to red, which implies the formation of Au NRs. The solution of the Au NRs was stable for more than two months. In Fig. 1(a), the high resolution transmission electron microscopic (TEM) image clearly shows the presence of the Au NRs and Fig. 1(b) shows their aspect ratio distribution with an average aspect ratio of 3.85.

The absorption spectra of the Au RNs dissolved in water were measured with a UV-visible spectrophotometer (UV-1700 Shimadzu) and one of them is illustrated in Figure 2(a). As shown in Fig. 2(a), it is evident that there are two SPR peaks. The first small SPR peak at 520 nm is due to the transverse mode perpendicular to the Au NRs; and the second SPR peak at 800 nm is due to the length area of the Au NRs which gives rise to the longitudinal mode of SPR with a higher peak.

The nonlinear optical properties of the Au NRs dissolved in water were investigated with femtosecond Z-scans at wavelengths around the peak position of longitudinal SPR. To minimize average power and reduce accumulative thermal effects, we employed 220-fs laser pulses at 1-kHz repetition rate. The laser pulses were generated by a mode-locked Ti: Sapphire laser (Quantronix, IMRA), which seeded a Ti: Sapphire regenerative amplifier (Quantronix, Titan). The wavelengths were tunable as the laser pulses passed through an optical parametric amplifier (Quantronix, TOPAS). The laser pulses were focused onto a 1-mm-thick quartz cuvette which contained the Au NR solution with a minimum beam waist of ~ 10 μm. By adding water to the Au NR solution, the linear transmittance of the solution was adjusted to be 70% at 800 nm. The incident and transmitted laser powers were monitored as the cuvette was moved (or Z-scanned) along the propagation direction of the laser pulses. Figure 2(b) displays, for example, typical open-aperture Z-scans carried out with the same irradiance of 0.5 $GW/cm^2$ at three different wavelengths, showing negative signs for the absorptive nonlinearities. We attribute this negativity to saturable absorption at the longitudinal mode of SPR.

To quantitatively determine the saturable absorption properties of the Au NR solution, we simulate the Z-scan data with a Z-scan theory[16], in which the nonlinear absorption is assumed to be $\alpha_2 I$, where $\alpha_2$ is the nonlinear absorption coefficient (which is proportional to the imaginary part of third-order nonlinear susceptibility, Im $\chi^{(3)}$), and $I$ is the light irradiance. From the best fit, we extracted the $\alpha_2$ value. Figure 2(b) shows the spectral dependence of the extracted $\alpha_2$ values, confirming the resonance effect. The largest $\alpha_2$ value (-1.5 cm/GW or Im $\chi^{(3)}$ = -1.2 x $10^{-12}$ *esu*) is observed at the peak position of longitudinal SPR. Schrof et al [17] reported the ratio of |Im $\chi^{(3)}$| to the linear

absorption coefficient ($\alpha$) to be ~ 3 x $10^{-14}$ *esu* cm at the SPR for 15-nm gold nanoparticles by using four-wave mixing technique with femtosecond laser pulses. By taking the linear absorption of the Au NR solution into consideration, we find that $|\text{Im}\ \chi^{(3)}/\alpha|$ is 4 x $10^{-13}$ *esu* cm, which is one order of magnitude greater than the ratio reported by Schrof et al [17] for gold nanoparticles.

It should be pointed out that the above results are obtained at a low excitation level, 0.5 GW/cm$^2$, which is far below the irradiances used to observe reverse saturable absorption discussed as follows. Fig. 3(a) displays the open-aperture Z-scans measured at higher excitation levels. It clearly illustrates that the nonlinear behavior alters from the saturable absorption (SA) to reverse saturable absorption (RSA) as the excitation intensity increases. The transformation from SA to RSA suggests that another nonlinear process takes place and becomes dominant. Such an interesting effect can be used for optical pulse compressor, optical switching and laser pulse narrowing.[18] The Z-scan data shows that, along with moving the Au NR solution towards the focus, the increase in the laser intensity induces bleaching in the ground-state plasmon absorption, which results in a transmittance increase (SA process). This process happens when the laser irradiance is less than ~ 7 GW/cm$^2$. As the laser intensity arises further (> 7 GW/cm$^2$), the RSA becomes dominant. In Fig. 3(b), the transmittance of the Au RN solution is plotted as a function of the laser irradiance, demonstrating that the transmittance decreases as the laser irradiance arises in the high-irradiance regime (>7 GW/cm$^2$). This nonlinear behavior is a typical characteristic for optical limiting phenomena. Such optical limiting is useful for applications of protecting optical sensors from intense laser pulses. If we define the limiting threshold as the incident irradiance, at which the solution transmittance falls to half of the linear transmittance, the limiting threshold of the Au NR solution is found to be ~ 24 GW/cm$^2$. The origin of the RSA is attributed to free-carrier absorption and/or the formation of strong light scattering centers due to the vaporization of the initial particles induced by the laser pulse.[9,10]

To evaluate the recovery time of the observed nonlinearities and to gain an insight of the underlying mechanism, we conducted a degenerate pump-probe experiment with 220-fs, 780-nm laser pulses, close to the peak position of longitudinal SPR (~ 800 nm). The inset of Fig. 3(b) displays the transient absorption signals of the Au NR solution. In this experiment, the pump irradiance was kept below 7 GW/cm$^2$, avoiding the RSA effect. The results in the inset of Fig. 3(b) confirm the SA processes measured using Z-scan as depicted in Fig. 3(a). In the pump-probe measurements with input irradiances less than 5 GW/cm$^2$, photo-excited electron dynamics in the Au NRs can be attributed to three relaxation processes: the excited electrons relax to the ground state through electron–electron ($\tau_{e\text{-}e}$ = 220 fs), electron–phonon ($\tau_{e\text{-}ph}$ = 3.8 ps), and phonon–phonon ($\tau_{ph\text{-}ph}$ = 100 ps) interactions. These interactions induce the further excitation of the hot electrons around the SPR peak. Thus, the broadband transient absorption subsides in this duration and gives way to the complete recovery of the bleached spectrum. Such optical response is interpreted in terms of frequency shift and broadening of the SPR and is related to the changes of the Au NR dielectric function induced by fast perturbation of the electron distribution. It is interesting to note that our measurements are comparable to the relaxation times ($\tau_{e\text{-}ph}$ = 3.9 ps and $\tau_{ph\text{-}ph}$ = 100 ps) reported for bleaching recovery at the transverse SPR.[19] For a higher input irradiance (~6.9 GW/cm$^2$), the relaxation processes become complicated because the measurement can not be fitted with the relaxation model based on the above–discussed three recovery processes. In particular, near the zero delay, an oscillatory signal superimposes to the recovery processes. The origin of this oscillation is unclear and it may be attributed to interplay between the SA and RSA process since the pump irradiance used is considerably close to the threshold for occurrence of RSA revealed by our Z-scans.

In conclusion, longitudinal SPR nonlinearities have been observed in Au NRs with femotosecond laser pulses, and their connection with the conduction-band electron dynamics has been discussed. Such observations are conducted using high-sensitivity femtosecond pump-probe and Z-scan techniques in Au NRs diluted in water. By focusing nonlinear optical measurements near the longitudinal mode of the SPR peak, the spectral dependence of SA is determined with a recovery time of a few ten ps at low excitation irradiances. Due to further absorption of excited electrons and/or nonlinear scattering, RSA occurs at high excitation levels ( > 7 $GW/cm^2$). Such RSA makes Au NRs an ideal candidate for optical limiting applications.

## Figure captions:

**Figure 1.** (**a**) TEM images of the Au NRs. (**b**) For the longitudinal mode of SPR peak at 800 nm, we obtain the average aspect ratio of 3.85.

**Figure 2.** (**a**) Linear and nonlinear absorption coefficients of the Au NRs. (**b**) Open-aperture Z-scans of the Au NR solution measured at different wavelengths. The irradiance used is kept at 0.5 GW/cm$^2$. The solid lines are the best-fit curves calculated by using the Z-scan theory reported in Ref. [16].

**Figure 3.** (a) Z-scans of the Au NR solution carried out at 800 nm with 220-fs laser pulses. The top curves are shifted vertically for clear presentation. (b) Transmittance of the Au NR solution measured as a function of the input irradiance at 800 nm with 220-fs laser pulses. The inset in (b) is the measurements of transient transmission on the Au NR solution conducted at a wavelength of 780 nm with 220-fs laser pulses. The solid lines are the best fits that give decay times $\tau_{e\text{-}e}$ = 220 fs, $\tau_{e\text{-}ph}$ = 3.8 ps, and $\tau_{ph\text{-}ph}$ = 100 ps for input irradiances (*I*) less than of 5 GW/cm$^2$.

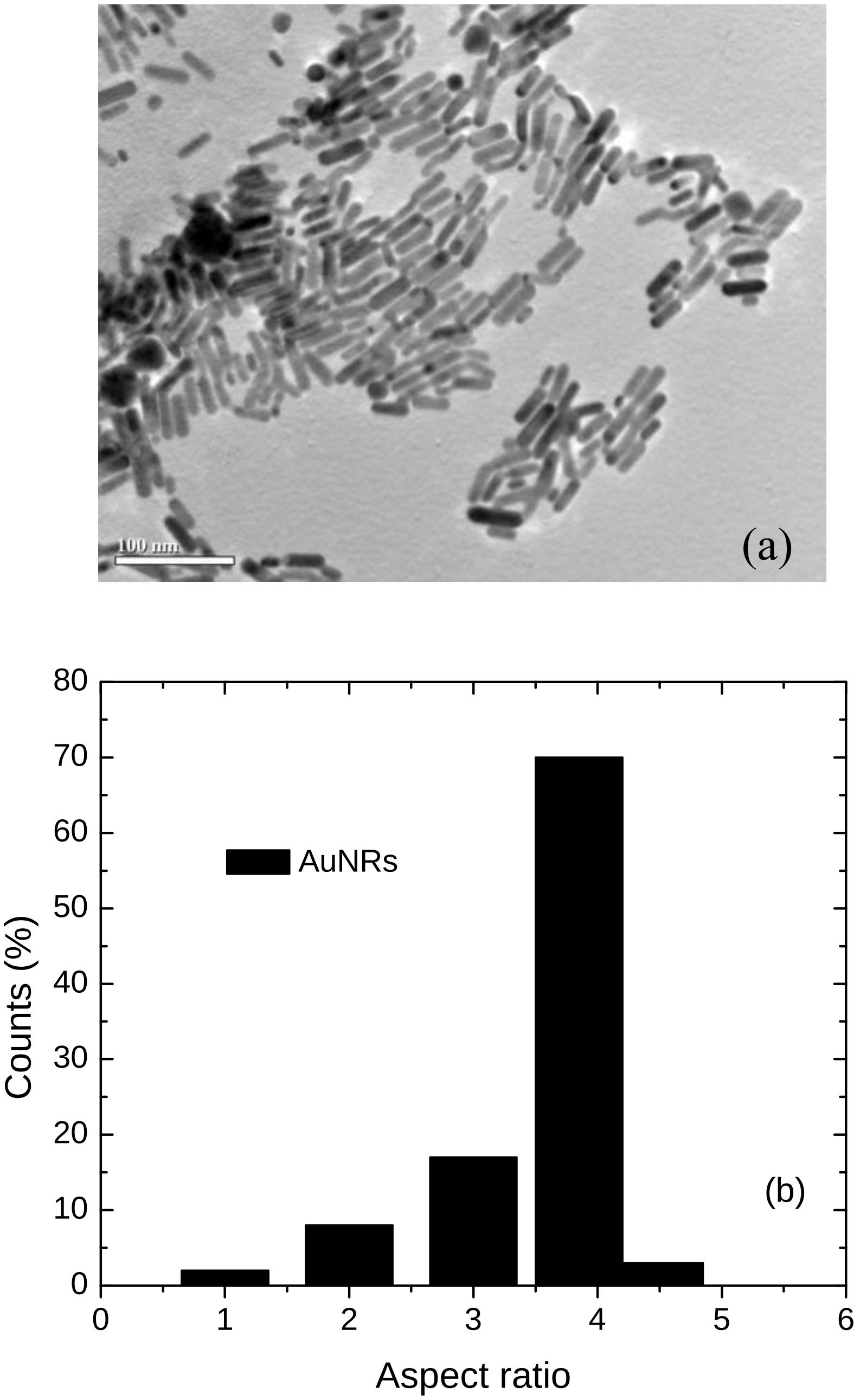

Figure 1.

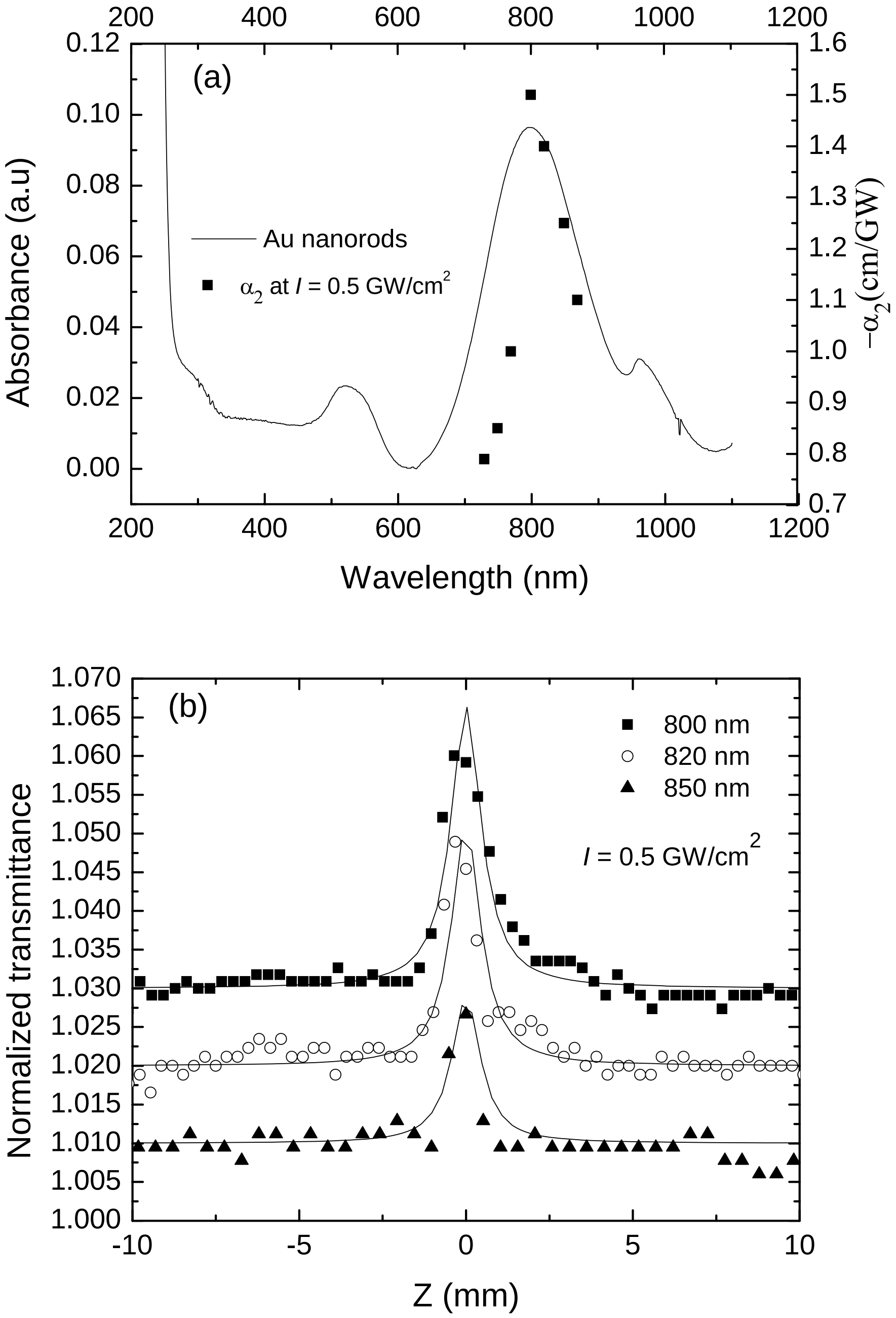

**Figure 2.**

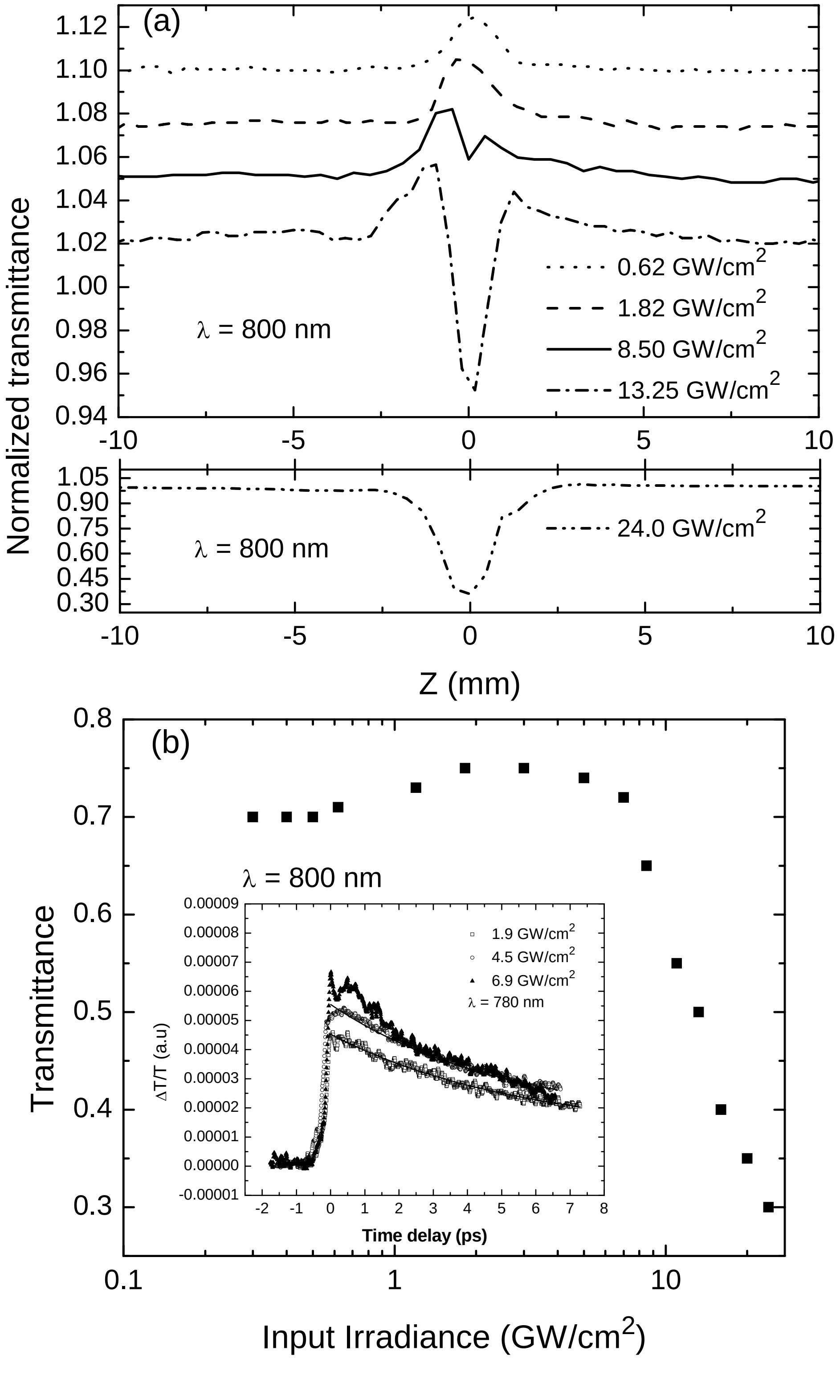

**Figure 3.**